\documentclass[letter]{aa}

\usepackage{graphicx}
\usepackage{txfonts}

\newcommand{\lya}{Ly$\alpha$}
\newcommand{\nv}{\ion{N}{v}$\lambda$1240}
\newcommand{\civ}{\ion{C}{iv}$\lambda$1549}
\newcommand{\heii}{\ion{He}{ii}$\lambda$1640}
\newcommand{\ciii}{\ion{C}{iii}]$\lambda$1909}

\begin{document}

\title{The mass-metallicity relation of high-z type-2 active galactic nuclei}

\author{K. Matsuoka\inst{1,2}
        \and
        T. Nagao\inst{3}
        \and
        A. Marconi\inst{1}
        \and
        R. Maiolino\inst{4,5}
        \and
        F. Mannucci\inst{2}
        \and
        G. Cresci\inst{2}
        \and
        K. Terao\inst{6}
        \and
        H. Ikeda\inst{7}
        }

\institute{Dipartimento di Fisica e Astronomia, Universit{\'a} degli Studi di Firenze,
           Via G. Sansone 1, I-50019 Sesto Fiorentino, Italy
           \and
           INAF -- Osservatorio Astrofisico di Arcetri,
           Largo Enrico Fermi 5, I-50125 Firenze, Italy\\
           \email{matsuoka@arcetri.astro.it}
           \and
           Research Center for Space and Cosmic Evolution,
           Ehime University, 2-5 Bunkyo-cho, Matsuyama 790-8577, Japan
           \and
           Cavendish Laboratory, University of Cambridge,
           19 J. J. Thomson Avenue, Cambridge CB3 0HE, UK
           \and
           Kavli Institute for Cosmology, University of Cambridge,
           Madingley Road, Cambridge CB3 0HA, UK
           \and
           Graduate School of Science and Engineering, Ehime University,
           2-5 Bunkyo-cho, Matsuyama 790-8577, Japan
           \and
           National Astronomical Observatory of Japan (NAOJ),
           2-21-1 Osawa, Mitaka, Tokyo 181-8588, Japan}

\date{}

\abstract{
The mass-metallicity relation (MZR) of type-2 active galactic nuclei (AGNs) at $1.2 < z < 4.0$ is investigated by using high-$z$ radio galaxies (HzRGs) and X-ray selected radio-quiet AGNs.
We combine new rest-frame ultraviolet (UV) spectra of two radio-quiet type-2 AGNs obtained with FOCAS on the Subaru Telescope with existing rest-frame UV emission lines, i.e., \civ, \heii, and \ciii, of a sample of 16 HzRGs and 6 additional X-ray selected type-2 AGNs, whose host stellar masses have been estimated in literature.
We divided our sample in three stellar mass bins and calculated averaged emission-line flux ratios of \civ/\heii\ and \ciii/\civ.
Comparing observed emission-line flux ratios with photoionization model predictions, we estimated narrow line region (NLR) metallicities for each mass bin.
We found that there is a positive correlation between NLR metallicities and stellar masses of type-2 AGNs at $z \sim 3$.
This is the first indication that AGN metallicities are related to their hosts, i.e., stellar mass.
Since NLR metallicities and stellar masses follow a similar relation as the MZR in star-forming galaxies at similar redshifts, our results indicate that NLR metallicities are related to those of the host galaxies.
This study highlights the importance of considering lower-mass X-ray selected AGNs in addition to radio galaxies to explore the metallicity properties of NLRs at high redshift.
}

\keywords{galaxies: active -- galaxies: evolution -- galaxies: nuclei -- quasars: emission lines -- quasars: general}

\maketitle

\section{Introduction}

Metallicity provides one of the most important clues in understanding galaxy formation and evolution since metal formation and enrichment are closely connected with the past star formation history in galaxies.
Generally, the metallicity of star-forming galaxies is obtained by using various emission lines at the rest-frame optical wavelengths (e.g., [\ion{O}{ii}]$\lambda$3727, H$\beta$, [\ion{O}{iii}]$\lambda$5007, H$\alpha$, and [\ion{N}{ii}]$\lambda$6584), emitted in \ion{H}{ii} regions \citep[e.g.,][]{2006A&A...459...85N,2017MNRAS.465.1384C}.
The relationship between mass of galaxies and their metal (the so-called mass-metallicity relation, MZR) has been investigated for about forty years \citep{1979A&A....80..155L}.
By using Sloan Digital Sky Survey (SDSS) imaging and spectroscopy of $\sim$ 53,000 star-forming galaxies at $z \sim 0.1$, \citet{2004ApJ...613..898T} made a major step forward in this field by obtaining a highly statistically sound relationship between stellar mass and gas-phase oxygen abundances in galaxies \citep[see also][]{2008ApJ...681.1183K,2009ApJ...691..140H,2013ApJ...765..140A}.
Furthermore, this correlation has been explored at high redshifts, up to $z \sim 3$, and a clear evolution with cosmological time, has been reported \citep[e.g.,][]{2005ApJ...635..260S,2008A&A...488..463M,2009MNRAS.398.1915M,2012MNRAS.421..262C,2013ApJ...763....9Y,2016MNRAS.463.2002H,2016ApJ...822...42O,2017ApJ...849...39S}.

The metallicity of star-forming galaxies at high redshift has been estimated through deep near-infrared spectroscopy.
However, since the emission lines for the metallicity diagnostics shift out of the $K$-band atmospheric window for galaxies at $z > 3.5$, studies are more challenging for higher-$z$ galaxies requiring space-based spectrograph.
An alternative approach to investigate the metallicity in the early universe is to focus on active galactic nuclei (AGNs) instead of star-forming galaxies.
Thanks to their high luminosity and strong emission lines in the wide wavelength range from the ultraviolet (UV) to the infrared, arising from gas clouds photoionized by their central accreting black hole, it is possible to measure the metallicity even in high-$z$ universe.
Since \citet{1992ApJ...391L..53H} proposed that broad-line region (BLR) metallicities ($Z_{\rm BLR}$) can be estimated from \nv, \civ, and \heii\ lines \citep[see also][]{1993ApJ...418...11H,1999ARA&A..37..487H}, AGN metallicities have been studied extensively.
By measuring emission-line flux ratios of SDSS quasars, i.e., \nv/\civ, (\ion{Si}{iv}$\lambda$1398+\ion{O}{iv}$\lambda$1402)/\civ, \ion{Al}{iii}$\lambda$1857/\civ, and \nv/\heii, \citet{2006A&A...447..157N} investigated BLR metallicities and their properties.
Based on a spectral stacking analysis, they found a positive correlation between BLR metallicities and AGN luminosities ($L_{\rm AGN}$): more luminous quasars show more metal-rich BLR clouds \citep[see also][]{1993ApJ...418...11H,2009A&A...494L..25J}.
By using SDSS spectra of 2383 quasars at $2.3 < z < 3.0$, \citet{2011A&A...527A.100M} revealed that the observed $L_{\rm AGN}$-$Z_{\rm BLR}$ trend is mostly a consequence of the relation between black-hole mass ($M_{\rm BH}$) and BLR metallicity.
This result indicates that AGN host galaxies may follow a MZR given the relation between black-hole mass and galaxy mass.
However, it is unclear if the BLR metallicities trace chemical properties of their host galaxies, since AGN broad lines originate from a very small region in galactic nuclei, $R_{\rm BLR} < 1$ pc \citep[e.g.,][]{2006ApJ...639...46S}, which may have evolved more rapidly and may not be representative of the metallicity in their host galaxies.

To investigate the metallicity on the scale of AGN host galaxies, the narrow line region (NLR) has been studied, since the typical NLR size is comparable to the size of their host galaxies, $R_{\rm NLR} \sim 10^{1-4}$ pc \citep[e.g.,][]{2006A&A...459...55B,2006A&A...456..953B}.
Furthermore, NLR masses are usually $\sim 10^{4-8} M_\odot$, much larger than those of the BLR, $\sim 10^{2-4} M_\odot$ \citep[e.g.,][]{2003ApJ...582..590B}.
Thus, NLR metallicities allow us to investigate chemical properties at host galaxy scales.
However, since it has been difficult to discover type-2 quasars in the high-$z$ universe, the NLR metallicity at $z > 1$ have been investigated only with high-$z$ radio galaxies (HzRGs) in previous studies.
\citet{2006A&A...447..863N} studied NLR metallicities of 51 HzRGs at $1.2 < z < 3.8$ with a metallicity diagnostic diagram involving \civ, \heii, and \ciii.
They reported that HzRGs do not show any redshift evolution of NLR metallicity: a lack of redshift evolution is similar to that seen for the BLR metallicities \citep{2006A&A...447..157N,2009A&A...494L..25J}.
\citet{2009A&A...503..721M} confirmed the absence of any significant NLR metallicity evolution up to $z \sim 4$ by observing nine HzRGs at $z > 2.7$.
Moreover, \citet{2011A&A...532L..10M} found that the most distant radio galaxy, TN J0924$-$2201 at $z = 5.19$, has already experienced significant chemical evolution.

On the other hand, \citet{2009A&A...503..721M} also found a positive correlation between NLR metallicities and AGN luminosities, as for the BLR metallicities.
This may be originated from the galaxy mass-metallicity relation, if we assume $L_{\rm AGN} \propto M_{\rm BH} \propto M_{\rm host}$.
However, there is still no direct evidence that NLR metallicities are related to their host properties, and in particular with their stellar mass ($M_\star$).
In this letter, we focus on HzRGs and X-ray selected type-2 AGNs at $1.2 < z < 4.0$ and investigate their NLR metallicities, adopting a diagnostic diagram of \civ, \heii, and \ciii\ emission lines.
Thus, by collecting their stellar masses from the recent literature we examine their dependence on NLR metallicities.

\begin{figure} 
\centering
\includegraphics[width=9cm]{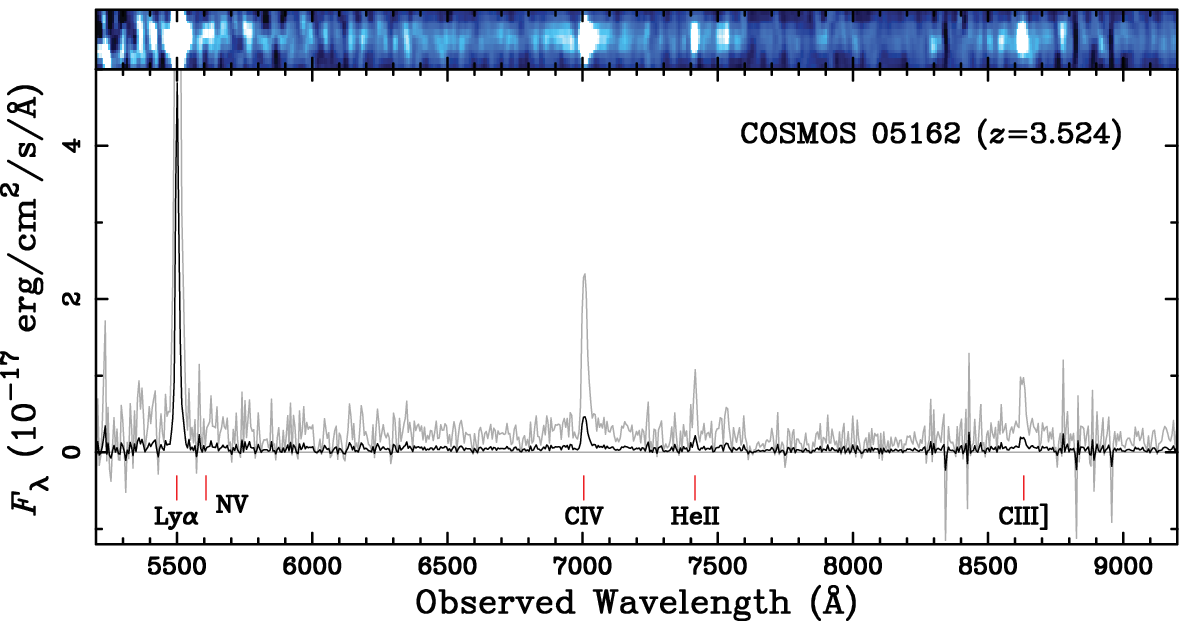}\vspace{3mm}
\includegraphics[width=9cm]{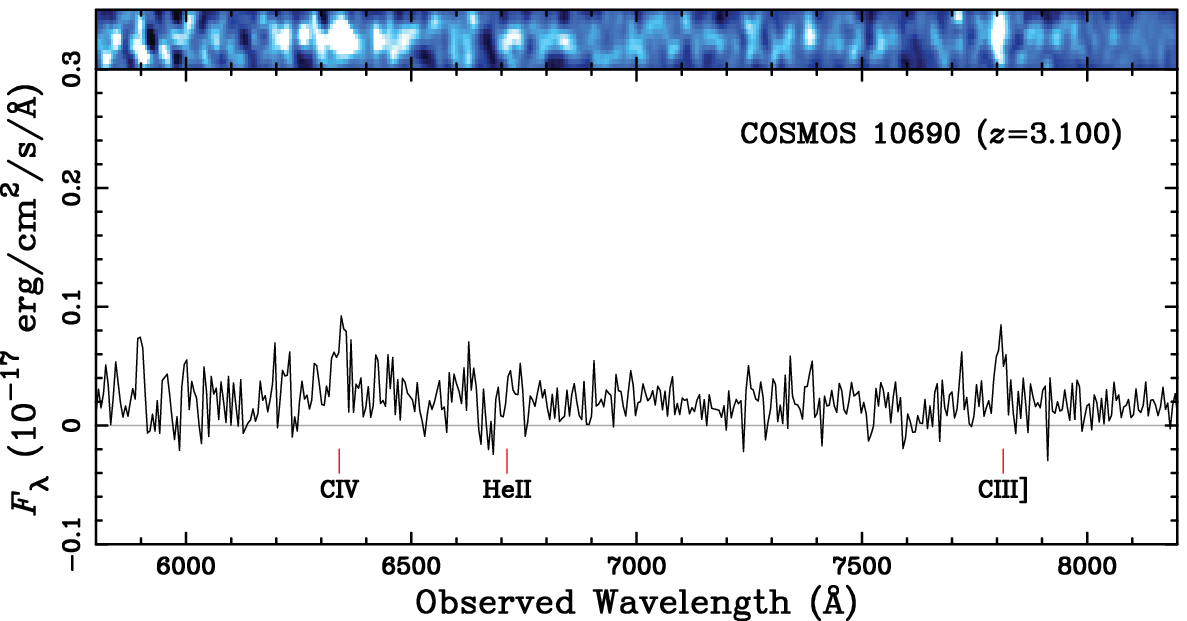}
\caption{Reduced spectra of two X-ray selected type-2 AGNs in the COSMOS field (ID: 05162 and 10690) obtained with FOCAS, after adopting 4-pixel binning in wavelength direction. Their two-dimensional spectra with a 2\farcs1 section are shown at the top, respectively. Red vertical lines indicate the central wavelengths of rest-frame UV emission lines, i.e., Ly$\alpha$, \nv, \civ, \heii, and \ciii. The grey line in the {\it top} panel is an arbitrary spectrum with five times the flux for recognising relatively weak emission lines.}
\label{fig:1}
\end{figure}

\section{Observations and Data from the Literature}

In order to study NLR metallicities, we utilize type-2 AGNs, i.e. in which the BLR and strong continuum emission are obscured allowing us accurate measurements of narrow-line emission.
First, we focus on HzRG samples \citep[e.g.,][]{2000A&A...362..519D} which have been identified as type-2 AGNs at high redshift with radio observations: e.g., ultra-steep-spectrum radio sources with radio spectral index $\alpha < -1.30$ ($S_\nu \propto \nu^\alpha$) are adopted as efficient HzRG tracers in \citet{2000A&A...362..519D} and over 160 sources are identified as HzRGs at $0 < z < 5.2$ with devoted optical spectroscopic observations.
Requiring that \civ, \heii, and \ciii\ lines are detected, 67 HzRGs at $1.2 < z < 4.0$ are collected from \citet{2006A&A...447..863N}, \citet{2007MNRAS.378..551B}, and \citet{2009A&A...503..721M,2011A&A...527A.100M}.
We listed these emission-line fluxes in Table~\ref{tab:1}.

As inferred from the tight correlation in the Hubble $K$-$z$ diagram \citep[e.g.,][]{2002AJ....123..637D}, HzRGs are usually associated with very massive elliptical galaxies \citep[$M_\star > 10^{11} M_\odot$; e.g.,][]{2007ApJS..171..353S}.
In order to also sample the lower stellar mass range, in this study we also focus on X-ray selected type-2 AGNs which have been discovered in recent observations.
First, we obtained optical spectra of two X-ray selected radio-quiet type-2 AGNs at $z > 3.0$ in the Cosmic Evolution Survey (COSMOS) field \citep{2011A&A...535A..80M} with FOCAS \citep[the Faint Object Camera And Spectrograph;][]{2002PASJ...54..819K} at the Subaru Telescope (24$-$25 December 2013).
Observations were performed with the 300R dispersion element and SY47 filter to cover the wavelength range of $\lambda_{\rm obs} = 4900-9500$\AA.
We used a slit width of 0\farcs6.
Standard data reduction procedures were applied to these spectra by using IRAF tasks.
Figure~\ref{fig:1} shows the final reduced spectra.
We measured emission-line fluxes of detected lines by fitting them with a Gaussian function using the IRAF task {\tt splot} (see Table~\ref{tab:1}).
For undetected lines, 3$\sigma$ upper limits are given by assuming an averaged line width of 25\AA\ calculated from the detected lines.
Then, we collected emission-line fluxes of 10 X-ray selected AGNs at $1.5 < z < 4.0$ from \citet{2006A&A...447..863N}.
Nine of them are X-ray selected type-2 AGNs in the Chandra Deep Field-South (CDF-S) field \citep{2004ApJS..155..271S} and the remaining CXO 52 is a type-2 quasar in the Lynx field identified as a hard X-ray source with {\it Chandra} \citep{2002ApJ...568...71S}.
Emission line fluxes of these 13 X-ray type-2 AGNs are listed in Table~\ref{tab:1}.

\begin{figure*}
\centering
\includegraphics[width=5.7cm]{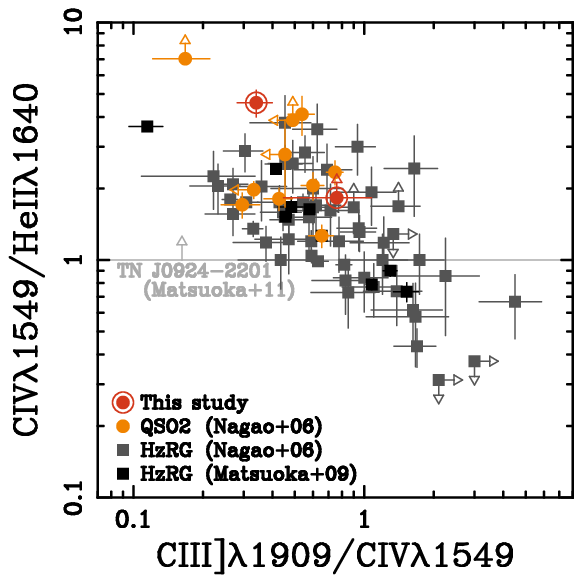}\hspace{5mm}
\includegraphics[width=5.7cm]{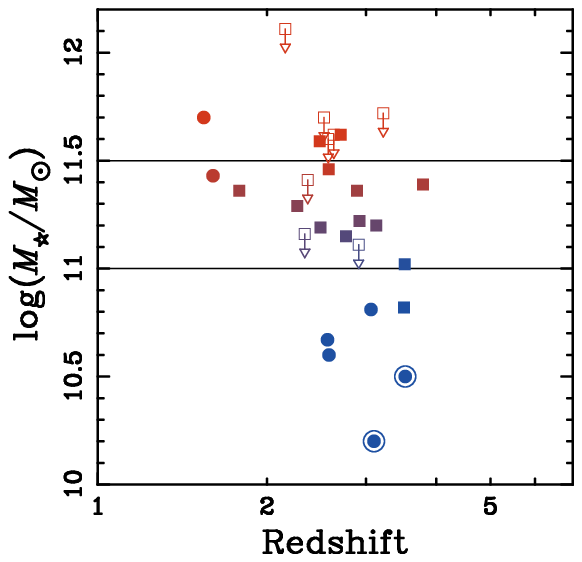}\hspace{5mm}
\includegraphics[width=5.7cm]{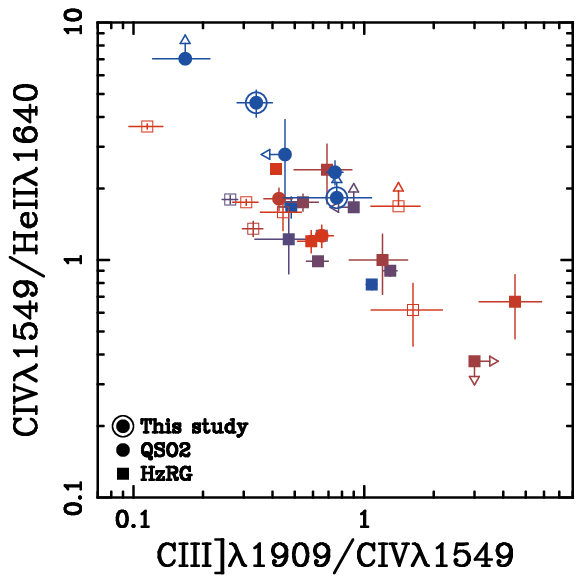}
\caption{Emission-line flux ratios of high-$z$ type-2 AGNs plotted on the diagnostic diagram of \civ/\heii\ versus \ciii/\civ\ ({\it left}). Gray- and black-filled squares and arrows denote HzRGs compiled by \citet{2006A&A...447..863N} and \citet{2009A&A...503..721M}, respectively. The \civ/\heii\ lower limit for TN J0924$-$2201 is shown as a light-gray arrow \citep{2011A&A...527A.100M}. Our new type-2 AGNs are shown as red-filled double circles (COSMOS 05162 and 10690). Orange circles denote X-ray selected type-2 AGNs in \citet{2006A&A...447..863N}. The {\it middle} panel shows the objects in our sample having an estimate of stellar mass as a function of redshift, and their emission-line flux ratios are plotted in the {\it right} panel. In the {\it middle} and {\it right} panels, symbols are same as the {\it left} panel but color-coded according to the stellar mass: bluer and redder for lower and higher mass objects, respectively. Note that open squares in these panels indicate HzRGs with upper limits of stellar masses.}
\label{fig:2}
\end{figure*}

We collected their stellar masses from the literature.
Regarding HzRGs, we matched our sample with stellar mass catalogs in \citet{2007ApJS..171..353S} and \citet{2010ApJ...725...36D}.
Overall, we obtained stellar masses of 21 HzRGs including 8 upper limits.
These studies derived stellar masses through a spectral energy distribution (SED) fit based on the {\it Spitzer} infrared (3.6--160 $\mu$m) data, covering the rest-frame optical to infrared wavelengths \citep[see][for more detail]{2007ApJS..171..353S}: \citet{2010ApJ...725...36D} followed the procedures of \citet{2007ApJS..171..353S} with the same method and photometric bands.
For X-ray selected AGNs in the COSMOS field, stellar masses are collected from \citet{2011A&A...535A..80M}.
These masses are also derived by SED fitting with multiwavelength data set of COSMOS.
For five additional AGNs in CDF-S, we adopted stellar masses derived using a tight correlation between rest-frame optical colors and stellar mass-to-light ratios in \citet{2010ApJ...720..368X}.
Altogether, we obtained stellar masses of 28 type-2 AGNs containing 8 HzRGs with upper limits (see Table~\ref{tab:1}).

\section{Result}

We calculate emission-line flux ratios of \civ/\heii\ and \ciii/\civ, that are sensitive to both the NLR gas metallicity and ionization parameter \citep[see][]{2006A&A...447..863N,2009A&A...503..721M}.
Figure~\ref{fig:2} shows our sample plotted on a diagnostic diagram of \civ, \heii, and \ciii\ lines in the {\it left} panel.
Our new type-2 AGNs are located in the same area as the X-ray selected AGNs in \citet{2006A&A...447..863N}.
In the {\it right} panel, 28 type-2 AGNs for which stellar masses were derived are plotted, color-coded according to their masses: we gave the stellar mass distribution as a function of redshift in the {\it middle} panel.
Thanks to X-ray selected type-2 AGNs, we could cover the low-$M_\star$ range, i.e., $\log (M_\star/M_\odot) < 11$, which could not be explored with HzRGs.

\section{Discussion}

\begin{figure*}
\centering
\includegraphics[width=5.7cm]{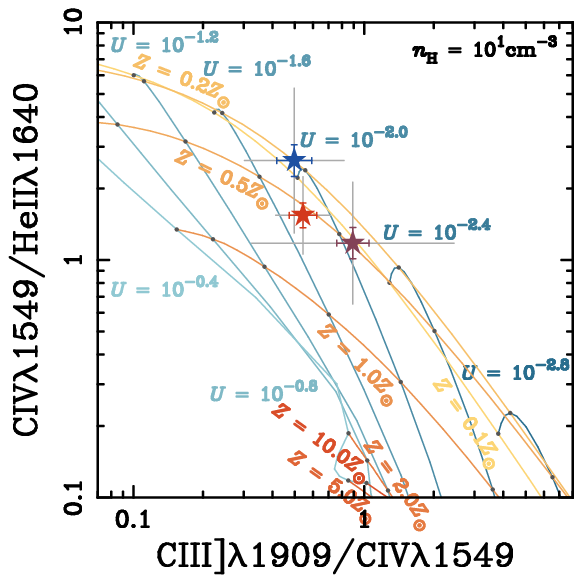}\hspace{5mm}
\includegraphics[width=5.7cm]{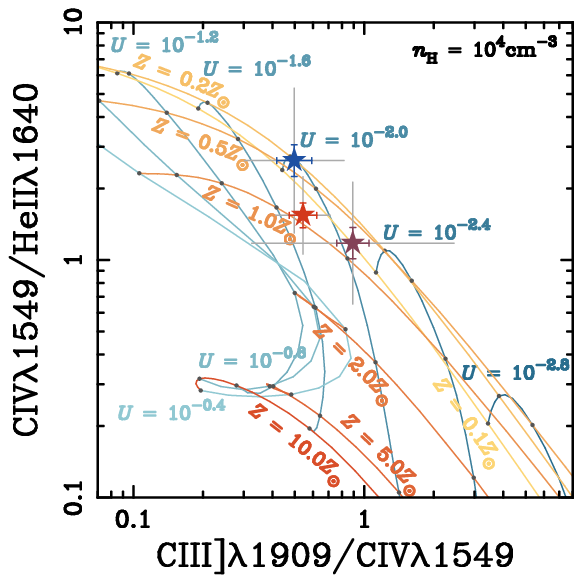}\hspace{5mm}
\includegraphics[width=5.7cm]{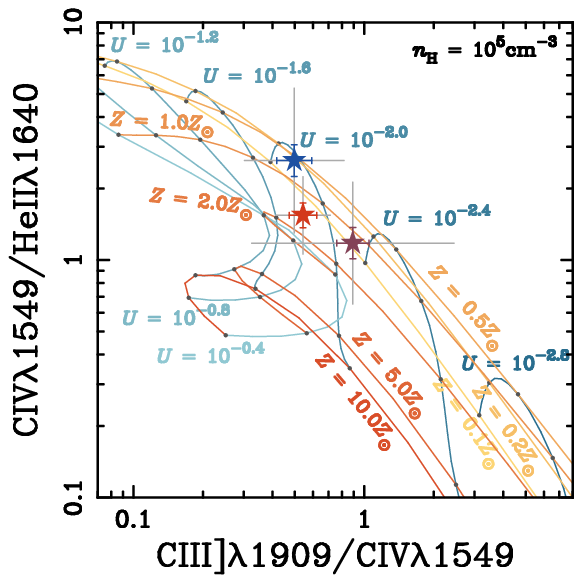}\vspace{3mm}
\includegraphics[width=5.7cm]{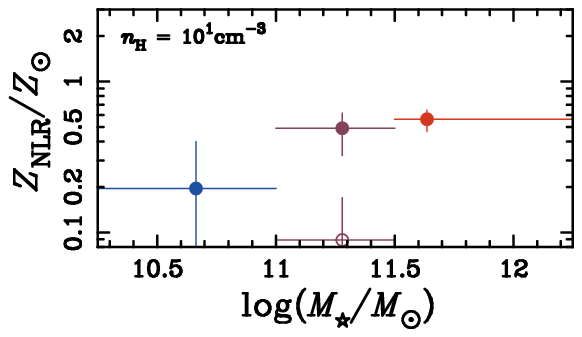}\hspace{5mm}
\includegraphics[width=5.7cm]{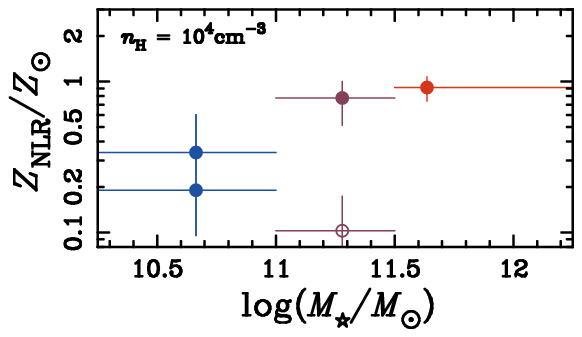}\hspace{5mm}
\includegraphics[width=5.7cm]{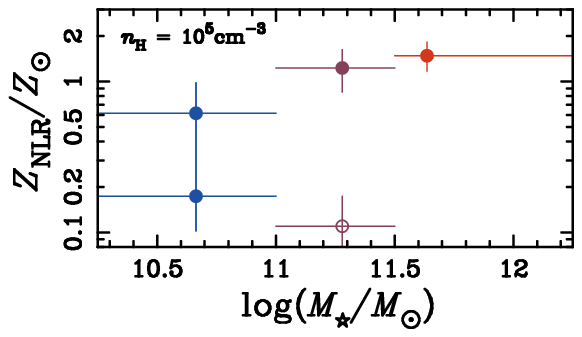}
\caption{Averaged emission-line flux ratios of low (blue), middle (purple), and high (red) mass subsamples plotted on the diagnostic diagram on \civ, \heii\ and \ciii\ ({\it top} pamels). The model predictions with different gas densities ($\log n_{\rm H} =$ 1, 4, and 5) are presented in the {\it left}, {\it middle}, and {\it right} panels, respectively. Constant metallicity and constant ionization parameter sequences are denoted by orange and blue solid lines, respectively. The gray bars denote the root mean square of the data distribution and the estimated errors in the averaged values are shown as color bars. The {\it bottom} panels show the NLR metallicities as a function of stellar masses, estimated with different three gas density models. Open circles would be less-likely solutions (see text).}
\label{fig:3}
\end{figure*}

In order to examine the relation between NLR metallicities and stellar masses statistically, we divided the 20 type-2 AGNs whose stellar masses are estimated without upper limits into three subsamples with stellar mass ranges, i.e., $\log (M_\star/M_\odot) < 11.0$, $11.0 < \log (M_\star/M_\odot) < 11.5$, and $\log (M_\star/M_\odot) > 11.5$ (these intervals are shown with horizontal lines in the {\it middle} panel in Figure~\ref{fig:2}).
Then, we calculated averaged emission-line flux ratios of \civ/\heii\ and \ciii/\civ\ for each subsample: we use 15 objects whose emission-line flux ratios are measured without upper or lower limits (i.e., three, nine, and three objects, respectively).
Note that our final sample would not be biased due to the discard of objects with upper or lower limits since the distribution of them show no obvious difference from the original sample (see the {\it right} panel of Figure~\ref{fig:2}).
Figure~\ref{fig:3} shows these averaged flux ratios on the diagnostic diagrams ({\it top} panels).
Using these averaged points, we estimate NLR metallicities for each stellar mass.

To estimate NLR metallicities from observational emission-line flux ratios, we carried out model calculations using the photoionization code Cloudy \citep[version 17.00;][]{1998PASP..110..761F,2017RMxAA..53..385F}: NLR clouds are mainly photoionized and not significantly affected by shocks when we focus on UV emission lines \citep[see][]{2009A&A...503..721M}.
We assumed three cloud models with hydrogen densities of $n_{\rm H} = 10^{1}$, 10$^{4}$, and 10$^{5}$ cm$^{-3}$.
For each model, we calculated emission-line flux ratios by spanning ionization parameter and metallicity in the following ranges $U = 10^{-3.0} - 10^{-0.4}$ and $Z_{\rm NLR} = 0.1 - 10 Z_\odot$, with step values of $\Delta \log U = 0.2$ and $\Delta \log (Z_{\rm NLR}/Z_\odot) = 0.01$, respectively (i.e., 8442 models in total).
We assume a typical AGN SED as photoionizing continuum radiation, using the ``table AGN'' command in Cloudy \citep{1987ApJ...323..456M}.
Chemical compositions of NLR gas clouds are scaled by maintaining solar abundance ratios except for helium and nitrogen.
We adopted analytical expressions for He and N relative abundances as a function of metallicities \citep[see][]{2000ApJ...542..224D,1992ApJ...384..508R,1998ApJ...497L...1V}.
We stopped calculations when the hydrogen ionization fraction dropped below 15\%.

Our model predictions of \civ/\heii\ and \ciii/\civ\ are overplotted on the diagnostic diagrams ({\it top} panels of Figure~\ref{fig:3}): metallicity-constant and $U$-constant sequences are denoted as orange and blue lines, respectively.
Our models confirm these photoionization models are useful to investigate NLR metallicities, although they show some dependences on hydrogen gas densities.
Comparing these model predictions with our $M_\star$-averaged line ratios we estimated mean NLR metallicities.
We show results of $Z_{\rm NLR}$ estimates as a function of stellar masses for each $n_{\rm H}$ model in bottom panels of Figure~\ref{fig:3}.
We estimated NLR metallicities for each $M_\star$-averaged line ratio and their errors from averaged emission-line flux ratios and their error bars, respectively, matching them to model calculations with a metallicity step of $\Delta \log (Z_{\rm NLR}/Z_\odot) = 0.01$.
Since absolute NLR metallicities are not determined uniquely due to dependences of gas densities, we compare the relative $Z_{\rm NLR}$ differences for each density.
This approach should be reliable since \citet{2016ApJ...816...23S} reported no dependence of gas density on stellar mass investigating the [\ion{O}{ii}]$\lambda\lambda$3726,3729 and/or [\ion{S}{ii}]$\lambda\lambda$6716,6731 doublets of local and high-$z$ star-forming galaxies at $z \sim 2.3$ \citep[see also][]{2015MNRAS.451.1284S,2017MNRAS.465.3220K}.
Note that in principle we derived two values of metallicities if emission-line flux ratios are located in a degenerated area on the model predictions.
However, lower metallicity in the middle-$M_\star$ bin is unplausible as a significant decrease of metallicity around medium-masses has never been seen for any population of galaxies, at any redshift, and not conceived by any model or cosmological simulation, therefore we exclude such lower metallicities solution in the middle-$M_\star$ bin, and consider only the upper solution for this bin.

As shown in these panels of Figure~\ref{fig:3}, we find positive correlations between NLR metallicities and stellar masses in type-2 AGNs at $1.2 < z < 4.0$ at different fixed densities.
This result is the first indication that AGN gas metallicities are related to their host properties, i.e., host galaxy masses in this case, most likely indicating $Z_{\rm NLR} \propto Z_{\rm gal}$ if we assume that AGN host galaxies have the same MZR in the star-forming galaxies.
We thus propose that NLR metallicity is correlated with host metallicity, although additional data sets would be needed to confirm this statistically.

The same averaged emission-line flux ratios shown in {\it top} panels of Figure~\ref{fig:3} may also be reproduced if there is an anti-correlation between stellar masses and densities instead of the MZR.
However, some observational studies of star-forming galaxies at $0 < z < 2.5$ indicated that electron densities do not show any such dependence on stellar masses.
For example, \citet{2017MNRAS.465.3220K} found no significant $M_\star$-$n_{\rm e}$ trend of star-forming galaxies at $z \sim 1.5$ although local SDSS galaxies indicate a weak positive correlation between electron density and stellar mass, indicating the opposite sense.
Moreover, \citet{2015MNRAS.451.1284S} reported any correlation between $M_\star$ and $n_{\rm e}$ cannot be identified with H$\alpha$-selected star-forming galaxies at $z = 2.5$ \citep[see also][]{2016ApJ...816...23S}.
Note here that we assume these $M_\star$-$n_{\rm e}$ relation of star-forming galaxies are applicable to NLR densities.
Therefore, we conclude that our mass-averaged emission-line flux ratios can be explained mainly by the dependence on NLR metallicities.

\begin{figure}
\centering
\includegraphics[width=8.6cm]{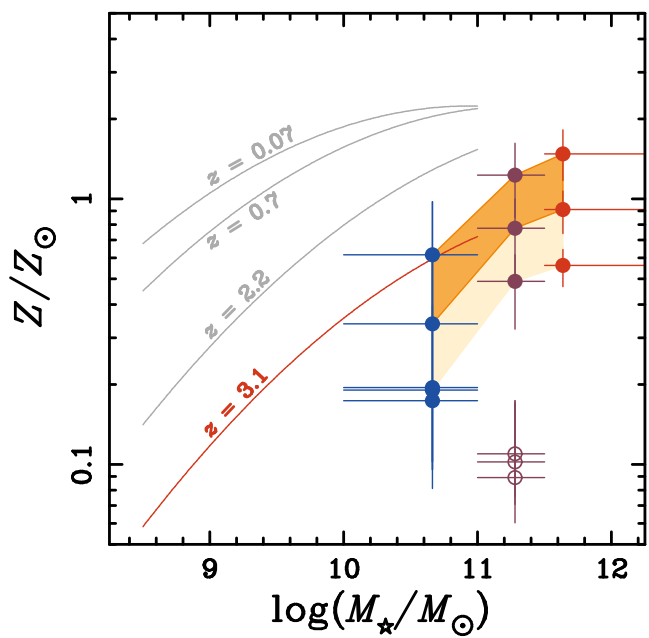}
\caption{The mass-metallicity relation of type-2 AGNs at $z \sim 3$ comparing with MZR of star-forming galaxies. Gray solid curves are MZR of star-forming galaxies at $z = 0.07$, 0.7, and 2.2, and red curve is for $z = 3.1$ galaxies in \citet{2009MNRAS.398.1915M}. Filled and open circles are same as Figure~\ref{fig:3}, and the orange area denotes MZR of type-2 AGNs which NLR metallicities are estimated with high-$n_{\rm H}$ models.}
\label{fig:4}
\end{figure}

Although there are uncertainties associated with density dependences, we compare our MZR of type-2 AGNs with those of galaxies.
Figure~\ref{fig:4} shows the redshift evolution of the MZRs for star-forming galaxies at $z = 0.07$, 0.7, 2.2, and 3.1 \citep{2008ApJ...681.1183K,2005ApJ...635..260S,2006ApJ...644..813E,2009MNRAS.398.1915M}, with our results are overplotted.
Since the mean redshift of our sample is $\sim 2.7$, we focus on the MZR of star-forming galaxies at $z = 3.1$ as a comparison.
As shown in this figure, we indicate that MZRs obtained with high-$n_{\rm H}$ models ($n_{\rm H} = 10^{4}$ and $10^{5}$ cm$^{-3}$ shown as an orange shaded area) seems consistent with MZR of star-forming galaxies.
Recently, Nitta et al. (in prep.) has found that NLR gas densities of high-$z$ AGNs are relatively higher than those of low-$z$ ones.
By investigating rest-frame optical emission lines of quasars, they estimated a typical gas density of $\sim 10^{5.5}$ cm$^{-3}$ \citep[see also][]{2012A&A...543A.143A}.

Note that some studies \citep[e.g.,][]{2006A&A...447..863N,2009A&A...503..721M} about the chemical evolution of HzRGs claimed that there is no significant metallicity evolution in NLRs, up to $z \sim 4$, apparently contradicting our result that HzRGs follow the redshift-dependent MZR of star-forming galaxies.
However, a possibility is that it is not possible to identify such redshift evolution only focusing on HzRGs hosted in massive elliptical galaxies.
Probably, the inclusion of X-ray selected AGNs with $\log (M_\star/M_\odot) \sim 10.5$ has helped to reveal such redshift-evolving MZR in type-2 AGNs as shown in this study.
Moreover, star formation rate (SFR) or gas mass should be crucial parameters to include, as recent studies have revealed that the relation between stellar mass, gas-phase metallicity, and SFR or gas mass (fundamental metallicity relation: FMR) is important to understand the chemical evolution of galaxies \citep[e.g.,][]{2010MNRAS.408.2115M,2010A&A...521L..53L,2016A&A...595A..48B}.
To investigate the FMR of type-2 AGNs statistically we need a larger sample and also access to SFR diagnostics that are not affected by the AGN activity, e.g., the far-infrared luminosity \citep[e.g.,][]{2006ApJ...642..694B,2015ApJ...807...28M,2017ApJ...838..128M}.

\section{Conclusion}

We have investigated the relation between NLR metallicities and stellar masses of type-2 AGNs at $1.2 < z < 4.0$ by using a pilot sample of 11 HzRGs and four X-ray selected type-2 AGNs and found the following results:
\begin{enumerate}
\item
We found indications of a positive correlation between NLR metallicities and stellar masses of type-2 AGNs at $1.2 < z < 4.0$; this is the first direct evidence that AGN metallicities are related to their host properties.
\item
Although there are uncertainties associated with the density dependence of the metallicity diagnostic diagrams, we have found that the mass-metallicity relation of type-2 AGNs, as inferred by us, are compatible with that of star-forming galaxies at similar redshifts, indicating that NLR metallicity is connected with host metallicities.
\end{enumerate}

\begin{table*}
\caption{Emission-line measurements of type-2 AGNs and their host stellar masses.}
\label{tab:1}
\centering
\begin{tabular}{lcrrrrrrr}
\hline\hline
Source           & $z$   & \multicolumn{5}{c}{Line Flux [$10^{-17}$ erg cm$^{-2}$ s$^{-1}$]}                       & $\log M_\star/M_\odot$ & Ref.\tablefootmark{a}\\
                 &            &  \lya & \nv   & \civ & \heii & \ciii & &\\
\hline\hline
\multicolumn{9}{c}{HzRGs in \citet{2006A&A...447..863N}}\\
\hline
USS 0003$-$019   & 1.541 &  --             & --              &   5.90$\pm$1.20 &   3.90$\pm$0.80 &   3.40$\pm$0.50 & --                     & --  \\
MG 0018$+$0940   & 1.586 &  --             & --              &   0.81$\pm$0.16 &   0.42$\pm$0.08 &   0.87$\pm$0.17 & --                     & --  \\
WN J0040$+$3857  & 2.606 &   4.10$\pm$0.82 & $<  0.20$       & $<  1.50$       & $<  0.30$       &   0.60$\pm$0.10 & --                     & --  \\
MG 0046$+$1102   & 1.813 &  --             & --              &   0.65$\pm$0.13 &   0.55$\pm$0.11 &   0.79$\pm$0.16 & --                     & --  \\
MG 0122$+$1923   & 1.595 & --              & --              &   0.32$\pm$0.06 &   0.38$\pm$0.08 &   0.32$\pm$0.06 & --                     & --  \\
USS 0200$+$015   & 2.229 &  17.40$\pm$2.20 & $<  0.40$       &   4.20$\pm$0.50 &   3.20$\pm$0.40 &   4.00$\pm$0.50 & --                     & --  \\
USS 0211$-$122   & 2.336 &   5.70$\pm$0.60 &   1.49$\pm$0.13 &   2.82$\pm$0.10 &   1.57$\pm$0.03 &   0.74$\pm$0.05 & $< 11.16$              & DB10\\
USS 0214$+$183   & 2.130 & --              & --              &   3.00$\pm$0.30 &   1.80$\pm$0.20 &   1.80$\pm$0.20 & --                     & --  \\
WN J0303$+$3733  & 2.504 &  10.30$\pm$2.06 & $<  0.20$       &   1.50$\pm$0.20 &   0.50$\pm$0.10 &   1.40$\pm$0.20 & --                     & --  \\
MG 0311$+$1532   & 1.986 & --              & --              &   0.34$\pm$0.07 &   0.20$\pm$0.04 &   0.21$\pm$0.04 & --                     & --  \\
BRL 0310$-$150   & 1.769 & --              & --              &  10.20$\pm$1.50 &   4.00$\pm$0.80 &   5.00$\pm$0.80 & --                     & --  \\
USS 0355$-$037   & 2.153 &  11.20$\pm$1.80 & $<  0.30$       &   2.70$\pm$0.60 &   3.70$\pm$0.70 &   2.30$\pm$0.50 & --                     & --  \\
USS 0448$+$091   & 2.037 &  12.20$\pm$1.40 & $<  0.40$       &   1.20$\pm$0.40 &   1.40$\pm$0.40 &   2.70$\pm$0.60 & --                     & --  \\
USS 0529$-$549   & 2.575 &   7.40$\pm$0.80 & $<  0.20$       &   0.40$\pm$0.10 &   0.60$\pm$0.10 &   1.80$\pm$0.30 & 11.46                  & Se07\\
4C $+$41.17      & 3.792 &  14.60$\pm$2.92 &   0.39$\pm$0.08 &   1.32$\pm$0.26 &   0.55$\pm$0.11 &   0.91$\pm$0.18 & 11.39                  & Se07\\
B3 0731$+$438    & 2.429 &  31.00$\pm$6.20 &   0.71$\pm$0.06 &   4.65$\pm$0.10 &   3.04$\pm$0.05 &   2.12$\pm$0.05 & --                     & --  \\
USS 0748$+$134   & 2.419 &   6.30$\pm$0.80 & $<  0.20$       &   1.80$\pm$0.30 &   1.50$\pm$0.30 &   1.40$\pm$0.30 & --                     & --  \\
USS 0828$+$193   & 2.572 &  13.30$\pm$1.30 &   3.67$\pm$0.59 &  18.17$\pm$0.45 &  10.38$\pm$0.14 &   5.59$\pm$0.68 & $< 11.60$              & DB10\\
BRL 0851$-$142   & 1.665 & --              & --              &   3.40$\pm$0.40 &   2.30$\pm$0.30 &   1.60$\pm$0.20 & --                     & --  \\
TN J0941$-$1628  & 1.644 & --              & --              &   3.20$\pm$0.50 &   0.90$\pm$0.20 &   2.00$\pm$0.30 & --                     & --  \\
USS 0943$-$242   & 2.923 &  20.10$\pm$2.00 &   1.12$\pm$0.06 &   4.65$\pm$0.11 &   4.71$\pm$0.13 &   2.93$\pm$0.31 & 11.22                  & Se07\\
MG 1019$+$0534   & 2.765 &   0.84$\pm$0.17 &   0.23$\pm$0.05 &   1.04$\pm$0.21 &   0.85$\pm$0.17 &   0.49$\pm$0.10 & 11.15                  & DB10\\
TN J1033$-$1339  & 2.427 &   9.80$\pm$1.96 & $<  0.10$       &   2.30$\pm$0.30 &   0.80$\pm$0.10 &   0.70$\pm$0.10 & --                     & --  \\
TN J1102$-$1651  & 2.111 &   2.70$\pm$0.54 & $<  0.20$       &   1.00$\pm$0.20 &   1.30$\pm$0.20 &   1.10$\pm$0.20 & --                     & --  \\
USS 1113$-$178   & 2.239 &   6.40$\pm$0.70 & $<  0.20$       &   1.70$\pm$0.40 &   0.70$\pm$0.20 &   2.80$\pm$0.30 & --                     & --  \\
3C 256           & 1.824 &  54.20$\pm$0.70 & $<  2.70$       &   5.23$\pm$0.08 &   5.47$\pm$0.05 &   4.28$\pm$0.33 & --                     & --  \\
USS 1138$-$262   & 2.156 &  13.90$\pm$1.60 & $<  0.30$       &   0.80$\pm$0.20 &   1.30$\pm$0.20 &   1.30$\pm$0.30 & $< 12.11$              & Se07\\
BRL 1140$-$114   & 1.935 & --              & --              &   1.00$\pm$0.30 &   0.50$\pm$0.10 &   0.60$\pm$0.10 & --                     & --  \\
4C $+$26.38      & 2.608 & --              & --              &   8.90$\pm$1.10 &   5.70$\pm$0.80 &   2.40$\pm$0.50 & --                     & --  \\
MG 1251$+$1104   & 2.322 &   2.31$\pm$0.46 & $<  0.06$       &   0.30$\pm$0.06 &   0.30$\pm$0.06 &   0.52$\pm$0.10 & --                     & --  \\
WN J1338$+$3532  & 2.769 &  17.80$\pm$3.56 & $<  0.60$       &   1.30$\pm$0.20 &   3.00$\pm$0.30 &   2.20$\pm$0.30 & --                     & --  \\
MG 1401$+$0921   & 2.093 & --              & --              &   0.41$\pm$0.08 &   0.50$\pm$0.10 &   0.34$\pm$0.07 & --                     & --  \\
3C 294           & 1.786 & 155.00$\pm$31.0 &   3.10$\pm$0.62 &  15.50$\pm$3.10 &  15.50$\pm$3.10 &  18.60$\pm$3.72 & 11.36                  & DB10\\
USS 1410$-$001   & 2.363 &  28.00$\pm$5.60 &   1.24$\pm$0.12 &   2.91$\pm$0.20 &   2.15$\pm$0.05 &   0.96$\pm$0.07 & $< 11.41$              & DB10\\
BRL 1422$-$297   & 1.632 & --              & --              &   4.30$\pm$0.60 &   2.10$\pm$0.40 &   1.00$\pm$0.20 & --                     & --  \\
USS 1425$-$148   & 2.349 &  20.70$\pm$4.14 & $<  0.70$       &   2.30$\pm$0.40 &   2.30$\pm$0.40 &   1.00$\pm$0.30 & --                     & --  \\
USS 1436$+$157   & 2.538 &  42.00$\pm$4.70 & $<  0.80$       &  17.00$\pm$1.90 &   6.00$\pm$0.80 &   9.40$\pm$1.70 & --                     & --  \\
3C 324           & 1.208 & --              & --              &   3.67$\pm$0.58 &   2.70$\pm$0.31 &   3.47$\pm$0.39 & --                     & --  \\
USS 1558$-$003   & 2.527 &  14.90$\pm$1.50 & $<  0.10$       &   2.70$\pm$0.30 &   1.70$\pm$0.20 &   1.20$\pm$0.20 & $< 11.70$              & DB10\\
BRL 1602$-$174   & 2.043 & --              & --              &  10.00$\pm$1.10 &   4.80$\pm$0.60 &   2.70$\pm$0.30 & --                     & --  \\
TXS J1650$+$0955 & 2.510 &  20.90$\pm$4.18 & $<  0.10$       &   3.20$\pm$0.40 &   2.70$\pm$0.30 &   1.20$\pm$0.30 & --                     & --  \\
8C 1803$+$661    & 1.610 & --              & --              &   5.30$\pm$1.06 &   2.60$\pm$0.52 &   1.90$\pm$0.38 & --                     & --  \\
4C 40.36         & 2.265 & --              &   2.91$\pm$0.56 &  13.61$\pm$0.51 &   7.77$\pm$0.56 &   7.39$\pm$1.18 & 11.29                  & Se07\\
BRL 1859$-$235   & 1.430 & --              & --              &   3.40$\pm$0.68 &   4.60$\pm$0.92 &   4.70$\pm$0.94 & --                     & --  \\
4C 48.48         & 2.343 & --              &   1.45$\pm$0.40 &   5.55$\pm$0.05 &   5.32$\pm$0.35 &   3.28$\pm$0.10 & --                     & --  \\
MRC 2025$-$218   & 2.630 &   4.00$\pm$0.80 &   0.62$\pm$0.21 &   0.69$\pm$0.07 & $<  0.41$       &   0.97$\pm$0.21 & $< 11.62$              & DB10\\
TXS J2036$+$0256 & 2.130 &   6.80$\pm$1.36 & $<  0.30$       & $<  0.90$       &   0.70$\pm$0.20 &   1.20$\pm$0.30 & --                     & --  \\
MRC 2104$-$242   & 2.491 & 57.00$\pm$11.40 & $<  3.80$       &   3.80$\pm$0.70 & $<  2.28$       & $<  3.42$       & 11.19                  & DB10\\
4C 23.56         & 2.483 &   8.00$\pm$1.60 &   1.15$\pm$0.13 &   1.80$\pm$0.15 &   1.50$\pm$0.10 &   1.06$\pm$0.10 & 11.59                  & Se07\\
MG 2109$+$0326   & 1.634 & --              & --              & $<  0.10$       &   0.32$\pm$0.06 &   0.21$\pm$0.04 & --                     & --  \\
MG 2121$+$1839   & 1.860 & --              & --              &   0.53$\pm$0.11 &   0.14$\pm$0.03 &   0.24$\pm$0.05 & --                     & --  \\
USS 2251$-$089   & 1.986 &   0.00$\pm$0.00 &   0.00$\pm$0.00 &   3.30$\pm$0.40 &   1.30$\pm$0.20 &   1.50$\pm$0.20 & --                     & --  \\
MG 2308$+$0336   & 2.457 &   2.93$\pm$0.59 &   0.57$\pm$0.11 &   0.63$\pm$0.13 &   0.39$\pm$0.08 &   0.45$\pm$0.09 & --                     & --  \\
4C $+$28.58      & 2.891 &   0.00$\pm$0.00 &   0.00$\pm$0.00 & $<  0.60$       &   1.60$\pm$0.40 &   1.80$\pm$0.40 & 11.36                  & DB10\\
\hline
\multicolumn{9}{c}{HzRGs in \citet{2007MNRAS.378..551B}}\\
\hline
MP J0340$-$6507  & 2.289 &   0.40$\pm$0.05 & --              &   0.15$\pm$0.05 &   0.26$\pm$0.05 &   0.25$\pm$0.05 & --                     & --  \\
TN J1941$-$1951  & 2.667 &   3.70$\pm$0.40 & --              &   0.90$\pm$0.10 &   0.40$\pm$0.10 &   0.20$\pm$0.10 & --                     & --  \\
MP J2352$-$6154  & 1.573 & --              & --              &   0.70$\pm$0.10 &   0.40$\pm$0.04 &   0.30$\pm$0.04 & --                     & --  \\
\hline
\end{tabular}
\end{table*}

\setcounter{table}{0}
\begin{table*}
\caption{Continued.}
\centering
\begin{tabular}{lcrrrrrrr}
\hline\hline
Source           & $z$   & \multicolumn{5}{c}{Line Flux [$10^{-17}$ erg cm$^{-2}$ s$^{-1}$]}                       & $\log M_\star/M_\odot$ & Ref.\tablefootmark{a}\\
                 &            &  \lya & \nv   & \civ & \heii & \ciii & &\\
\hline\hline
\multicolumn{9}{c}{HzRGs in \citet{2009A&A...503..721M}}\\
\hline
TN J0121$+$1320  & 3.517 &   1.58$\pm$0.01 & $<  0.09$       &   0.26$\pm$0.01 &   0.33$\pm$0.01 &   0.28$\pm$0.01 & 11.02                  & DB10\\
TN J0205$+$2242  & 3.507 &   8.93$\pm$0.02 & $<  0.10$       &   0.87$\pm$0.03 &   0.52$\pm$0.05 &   0.42$\pm$0.05 & 10.82                  & DB10\\
MRC 0316$-$257   & 3.130 &   2.02$\pm$0.01 & $<  0.04$       &   0.27$\pm$0.01 &   0.30$\pm$0.01 &   0.35$\pm$0.02 & 11.20                  & DB10\\
USS 0417$-$181   & 2.773 &   0.85$\pm$0.02 & $<  0.10$       &   0.36$\pm$0.03 &   0.49$\pm$0.02 &   0.55$\pm$0.05 & --                     & --  \\
TN J0920$-$0712  & 2.758 &  30.50$\pm$0.03 &   1.02$\pm$0.01 &   3.37$\pm$0.01 &   2.06$\pm$0.01 &   1.95$\pm$0.03 & --                     & --  \\
WN J1123$+$3141  & 3.221 &   1.88$\pm$0.01 &   1.70$\pm$0.01 &   1.57$\pm$0.01 &   0.43$\pm$0.01 &   0.18$\pm$0.03 & $< 11.72$              & DB10\\
4C $+$24.28      & 2.913 &   4.59$\pm$0.02 &   1.23$\pm$0.01 &   1.24$\pm$0.02 &   0.98$\pm$0.01 &   0.81$\pm$0.04 & $< 11.11$              & DB10\\
USS 1545$-$234   & 2.751 &   1.76$\pm$0.01 &   1.34$\pm$0.03 &   1.34$\pm$0.02 &   0.88$\pm$0.01 &   0.61$\pm$0.03 & --                     & --  \\
USS 2202$+$128   & 2.705 &   2.95$\pm$0.01 &   0.16$\pm$0.02 &   0.70$\pm$0.01 &   0.29$\pm$0.01 &   0.29$\pm$0.01 & 11.62                  & DB10\\
\hline\hline
\multicolumn{9}{c}{The most distant radio galaxy in \citet{2011A&A...532L..10M}}\\
\hline
TN J0924$-$2201  & 5.190 &  16.10$\pm$0.56 & $<  1.25$       &   5.46$\pm$0.52 & $<  5.54$       & --              & --                     & --  \\
\hline
\multicolumn{9}{c}{X-ray selected radio-quiet type-2 AGNs in this study}\\
\hline
COSMOS 05162     & 3.524 &  76.80$\pm$1.26 & $<  5.84$       &  11.90$\pm$0.77 &   2.59$\pm$0.29 &   4.05$\pm$0.65 & 10.50                  & Ma11\\
COSMOS 10690     & 3.100 &  $< 12.16$      & $<  6.82$       &   2.25$\pm$0.68 & $<  1.23$       &   1.71$\pm$0.48 & 10.20                  & Ma11\\
\hline
\multicolumn{9}{c}{X-ray selected type-2 AGNs in \citet{2006A&A...447..863N}}\\
\hline
CDFS$-$027       & 3.064 &  12.60$\pm$0.70 &   2.50$\pm$0.70 &   6.40$\pm$0.50 &   2.30$\pm$0.90 & $<  2.90$       & 10.81                  & Xu10\\
CDFS$-$031       & 1.603 & --              & --              &  24.10$\pm$1.40 &  13.30$\pm$1.20 &  10.30$\pm$1.30 & 11.43                  & Xu10\\
CDFS$-$057       & 2.562 & 112.20$\pm$1.30 &   8.40$\pm$1.40 &  17.80$\pm$0.80 &   7.60$\pm$0.80 &  13.30$\pm$0.90 & 10.67                  & Xu10\\
CDFS$-$112a      & 2.940 &  58.30$\pm$0.50 &  14.60$\pm$0.80 &  15.20$\pm$1.00 &   8.90$\pm$0.90 &   4.50$\pm$0.80 & --                     & --  \\
CDFS$-$153       & 1.536 & --              & --              &  25.50$\pm$1.40 &   6.20$\pm$1.10 &  13.70$\pm$1.60 & --                     & --  \\
CDFS$-$202       & 3.700 &  78.10$\pm$1.00 &  26.80$\pm$1.10 &  38.90$\pm$1.10 &  19.70$\pm$1.50 & $< 12.90$       & --                     & --  \\
CDFS$-$263b      & 3.660 &  70.90$\pm$0.70 &   4.60$\pm$0.70 &  15.50$\pm$0.80 & $<  4.00$       & $<  7.60$       & --                     & --  \\
CDFS$-$531       & 1.544 & --              & --              &  22.00$\pm$1.40 &  17.40$\pm$1.50 &  14.40$\pm$1.50 & 11.70                  & Xu10\\
CDFS$-$901       & 2.578 &  37.10$\pm$0.60 &   6.50$\pm$0.80 &  19.70$\pm$1.00 & $<  2.80$       &   3.30$\pm$0.90 & 10.60                  & Xu10\\
CXO 52           & 3.288 & 189.00$\pm$4.00 &   6.00$\pm$1.20 &  35.00$\pm$2.00 &  17.00$\pm$2.00 &  21.00$\pm$2.00 & --                     & --  \\
\hline
\end{tabular}
\tablefoot{
\tablefoottext{a}{DB10 $=$ \citet{2010ApJ...725...36D}; Se07 $=$ \citet{2007ApJS..171..353S}; Ma11 $=$ \citet{2011A&A...535A..80M}; Xu10 $=$ \citet{2010ApJ...720..368X}.}
}
\end{table*}

\begin{acknowledgements}
We would like to thank Takashi Hattori for assisting our FOCAS observations (S13B-019 and S14B-003).
We are also grateful to Gray J. Ferland for providing the great photoionization code Cloudy.
Data analysis were in part carried out on common-use data analysis computer system at the Astronomy Data Center, ADC, of the National Astronomical Observatory of Japan (NAOJ).
K.M. is supported by Japan Society for the Promotion of Science (JSPS) Overseas Research Fellowships.
T.N. is supported by JSPS KAKENHI Grant Nos. 16H01101, 16H03958, and 17H01114.
R.M. acknowledges support by the Science and Technology Facilities Council (STFC) and the ERC Advanced Grant 695671 ``QUENCH''.
\end{acknowledgements}


\begin{thebibliography}{}
\bibitem[Andrews \& Martini(2013)]{2013ApJ...765..140A} Andrews, B.~H., \& Martini, P.\ 2013, \apj, 765, 140
\bibitem[Araki et al.(2012)]{2012A&A...543A.143A} Araki, N., Nagao, T., Matsuoka, K., et al.\ 2012, \aap, 543, A143
\bibitem[Baldwin et al.(2003)]{2003ApJ...582..590B} Baldwin, J.~A., Ferland, G.~J., Korista, K.~T., et al.\ 2003, \apj, 582, 590
\bibitem[Beelen et al.(2006)]{2006ApJ...642..694B} Beelen, A., Cox, P., Benford, D.~J., et al.\ 2006, \apj, 642, 694
\bibitem[Bennert et al.(2006a)]{2006A&A...456..953B} Bennert, N., Jungwiert, B., Komossa, S., et al.\ 2006, \aap, 456, 953
\bibitem[Bennert et al.(2006b)]{2006A&A...459...55B} Bennert, N., Jungwiert, B., Komossa, S., et al.\ 2006, \aap, 459, 55
\bibitem[Bornancini et al.(2007)]{2007MNRAS.378..551B} Bornancini, C.~G., De Breuck, C., de Vries, W., et al.\ 2007, \mnras, 378, 551
\bibitem[Bothwell et al.(2016)]{2016A&A...595A..48B} Bothwell, M.~S., Maiolino, R., Cicone, C., et al.\ 2016, \aap, 595, A48
\bibitem[Curti et al.(2017)]{2017MNRAS.465.1384C} Curti, M., Cresci, G., Mannucci, F., et al.\ 2017, \mnras, 465, 1384
\bibitem[Cresci et al.(2012)]{2012MNRAS.421..262C} Cresci, G., Mannucci, F., Sommariva, V., et al.\ 2012, \mnras, 421, 262
\bibitem[De Breuck et al.(2000)]{2000A&A...362..519D} De Breuck, C., R{\"o}ttgering, H., Miley, G., et al.\ 2000, \aap, 362, 519
\bibitem[De Breuck et al.(2002)]{2002AJ....123..637D} De Breuck, C., van Breugel, W., Stanford, S.~A., et al.\ 2002, \aj, 123, 637
\bibitem[De Breuck et al.(2010)]{2010ApJ...725...36D} De Breuck, C., Seymour, N., Stern, D., et al.\ 2010, \apj, 725, 36
\bibitem[Dopita et al.(2000)]{2000ApJ...542..224D} Dopita, M.~A., Kewley, L.~J., Heisler, C.~A., \& Sutherland, R.~S.\ 2000, \apj, 542, 224
\bibitem[Erb et al.(2006)]{2006ApJ...644..813E} Erb, D.~K., Shapley, A.~E., Pettini, M., et al.\ 2006, \apj, 644, 813
\bibitem[Ferland et al.(1998)]{1998PASP..110..761F} Ferland, G.~J., Korista, K.~T., Verner, D.~A., et al.\ 1998, \pasp, 110, 761
\bibitem[Ferland et al.(2017)]{2017RMxAA..53..385F} Ferland, G.~J., Chatzikos, M., Guzm{\'a}n, F., et al.\ 2017, \rmxaa, 53, 385
\bibitem[Hamann \& Ferland(1992)]{1992ApJ...391L..53H} Hamann, F., \& Ferland, G.\ 1992, \apjl, 391, L53
\bibitem[Hamann \& Ferland(1993)]{1993ApJ...418...11H} Hamann, F., \& Ferland, G.\ 1993, \apj, 418, 11
\bibitem[Hamann \& Ferland(1999)]{1999ARA&A..37..487H} Hamann, F., \& Ferland, G.\ 1999, \araa, 37, 487
\bibitem[Hayashi et al.(2009)]{2009ApJ...691..140H} Hayashi, M., Motohara, K., Shimasaku, K., et al.\ 2009, \apj, 691, 140
\bibitem[Hunt et al.(2016)]{2016MNRAS.463.2002H} Hunt, L., Dayal, P., Magrini, L., \& Ferrara, A.\ 2016, \mnras, 463, 2002
\bibitem[Jiang et al.(2007)]{2007AJ....134.1150J} Jiang, L., Fan, X., Vestergaard, M., et al.\ 2007, \aj, 134, 1150
\bibitem[Juarez et al.(2009)]{2009A&A...494L..25J} Juarez, Y., Maiolino, R., Mujica, R., et al.\ 2009, \aap, 494, L25
\bibitem[Kaasinen et al.(2017)]{2017MNRAS.465.3220K} Kaasinen, M., Bian, F., Groves, B., Kewley, L.~J., \& Gupta, A.\ 2017, \mnras, 465, 3220
\bibitem[Kashikawa et al.(2002)]{2002PASJ...54..819K} Kashikawa, N., Aoki, K., Asai, R., et al.\ 2002, \pasj, 54, 819
\bibitem[Kewley \& Ellison(2008)]{2008ApJ...681.1183K} Kewley, L.~J., \& Ellison, S.~L.\ 2008, \apj, 681, 1183-1204
\bibitem[Lara-L{\'o}pez et al.(2010)]{2010A&A...521L..53L} Lara-L{\'o}pez, M.~A., Cepa, J., Bongiovanni, A., et al.\ 2010, \aap, 521, L53
\bibitem[Lequeux et al.(1979)]{1979A&A....80..155L} Lequeux, J., Peimbert, M., Rayo, J.~F., et al.\ 1979, \aap, 80, 155
\bibitem[Mainieri et al.(2011)]{2011A&A...535A..80M} Mainieri, V., Bongiorno, A., Merloni, A., et al.\ 2011, \aap, 535, A80
\bibitem[Maiolino et al.(2008)]{2008A&A...488..463M} Maiolino, R., Nagao, T., Grazian, A., et al.\ 2008, \aap, 488, 463
\bibitem[Mannucci et al.(2009)]{2009MNRAS.398.1915M} Mannucci, F., Cresci, G., Maiolino, R., et al.\ 2009, \mnras, 398, 1915
\bibitem[Mannucci et al.(2010)]{2010MNRAS.408.2115M} Mannucci, F., Cresci, G., Maiolino, R., et al.\ 2010, \mnras, 408, 2115
\bibitem[Mathews \& Ferland(1987)]{1987ApJ...323..456M} Mathews, W.~G., \& Ferland, G.~J.\ 1987, \apj, 323, 456
\bibitem[Matsuoka et al.(2009)]{2009A&A...503..721M} Matsuoka, K., Nagao, T., Maiolino, R., et al.\ 2009, \aap, 503, 721
\bibitem[Matsuoka et al.(2011)]{2011A&A...527A.100M} Matsuoka, K., Nagao, T., Marconi, A., et al.\ 2011, \aap, 527, A100
\bibitem[Matsuoka et al.(2011)]{2011A&A...532L..10M} Matsuoka, K., Nagao, T., Maiolino, R., et al.\ 2011, \aap, 532, L10
\bibitem[Matsuoka \& Woo(2015)]{2015ApJ...807...28M} Matsuoka, K., \& Woo, J.-H.\ 2015, \apj, 807, 28
\bibitem[Matsuoka \& Ueda(2017)]{2017ApJ...838..128M} Matsuoka, K., \& Ueda, Y.\ 2017, \apj, 838, 128
\bibitem[Nagao et al.(2006b)]{2006A&A...447..157N} Nagao, T., Marconi, A., \& Maiolino, R.\ 2006, \aap, 447, 157
\bibitem[Nagao et al.(2006c)]{2006A&A...447..863N} Nagao, T., Maiolino, R., \& Marconi, A.\ 2006, \aap, 447, 863
\bibitem[Nagao et al.(2006a)]{2006A&A...459...85N} Nagao, T., Maiolino, R., \& Marconi, A.\ 2006, \aap, 459, 85
\bibitem[Onodera et al.(2016)]{2016ApJ...822...42O} Onodera, M., Carollo, C.~M., Lilly, S., et al.\ 2016, \apj, 822, 42
\bibitem[Russell \& Dopita(1992)]{1992ApJ...384..508R} Russell, S.~C., \& Dopita, M.~A.\ 1992, \apj, 384, 508
\bibitem[Sanders et al.(2016)]{2016ApJ...816...23S} Sanders, R.~L., Shapley, A.~E., Kriek, M., et al.\ 2016, \apj, 816, 23
\bibitem[Savaglio et al.(2005)]{2005ApJ...635..260S} Savaglio, S., Glazebrook, K., Le Borgne, D., et al.\ 2005, \apj, 635, 260
\bibitem[Seymour et al.(2007)]{2007ApJS..171..353S} Seymour, N., Stern, D., De Breuck, C., et al.\ 2007, \apjs, 171, 353
\bibitem[Shimakawa et al.(2015)]{2015MNRAS.451.1284S} Shimakawa, R., Kodama, T., Steidel, C.~C., et al.\ 2015, \mnras, 451, 1284
\bibitem[Stern et al.(2002)]{2002ApJ...568...71S} Stern, D., Moran, E.~C., Coil, A.~L., et al.\ 2002, \apj, 568, 71
\bibitem[Suganuma et al.(2006)]{2006ApJ...639...46S} Suganuma, M., Yoshii, Y., Kobayashi, Y., et al.\ 2006, \apj, 639, 46
\bibitem[Suzuki et al.(2017)]{2017ApJ...849...39S} Suzuki, T.~L., Kodama, T., Onodera, M., et al.\ 2017, \apj, 849, 39
\bibitem[Szokoly et al.(2004)]{2004ApJS..155..271S} Szokoly, G.~P., Bergeron, J., Hasinger, G., et al.\ 2004, \apjs, 155, 271
\bibitem[Tremonti et al.(2004)]{2004ApJ...613..898T} Tremonti, C.~A., Heckman, T.~M., Kauffmann, G., et al.\ 2004, \apj, 613, 898
\bibitem[van Zee et al.(1998)]{1998ApJ...497L...1V} van Zee, L., Salzer, J.~J., \& Haynes, M.~P.\ 1998, \apjl, 497, L1
\bibitem[Xue et al.(2010)]{2010ApJ...720..368X} Xue, Y.~Q., Brandt, W.~N., Luo, B., et al.\ 2010, \apj, 720, 368
\bibitem[Yuan et al.(2013)]{2013ApJ...763....9Y} Yuan, T.-T., Kewley, L.~J., \& Richard, J.\ 2013, \apj, 763, 9
\end{thebibliography}
\end{document}